\documentstyle[preprint,aps]{revtex}
\draft

\begin{document}
\title{\bf $ $ \\
Phase--coherence Effects in Antidot Lattices:\\
A Semiclassical Approach to Bulk Conductivity}

\author{Klaus Richter\ddag \\
$ $ }

\address{Division de Physique Th\'eorique \footnote{Unit\'e de
Recherche des Universit\'es Paris XI et Paris VI associ\'ee au CNRS},
Institut de Physique Nucl\'eaire, F-91406 Orsay Cedex, France }

\date{September 23, 1994}
\maketitle
\begin{abstract}
We derive semiclassical expressions for the Kubo conductivity tensor.
Within our approach the oscillatory parts of the diagonal and
Hall conductivity are given as sums over contributions from classical
periodic orbits in close relation to Gutzwiller's trace formula for
the density of states. Taking into account the effects of weak disorder
and temperature we reproduce recently observed anomalous phase coherence
oscillations in the conductivity of large antidot arrays.
\end{abstract}

\narrowtext
\newpage
Interference phenomena in quantum electron transport through small
microstructures are usually interpreted within two complementary frameworks:
The Landauer--B\"uttiker formalism commonly describes transmission through
single phase coherent devices \cite{Been91}.
The current is expressed in terms of (a sum
over) conductance coefficients between channels in different leads
attached. On the other hand linear response theory (Kubo formalism) has
proved useful to treat bulk transport properties of samples with a size
exceeding the phase breaking length.
Besides coherence effects related to the presence of disorder (e.g.,
universal conductance fluctuations, weak localization) the development
of high mobility devices has opened experimental access to the ballistic limit
where the elastic mean free path is considerably larger than the system size
and the conductance reflects the geometry or potential landscape of the
microstructures. This has especially oriented interest to questions how
classically regular or chaotic electron dynamics manifests itself on the level
of quantum transport \cite{Marc92,Wei93,Rodol90}.
In this spirit a semiclassical approach to conductance within the
Landauer--B\"uttiker framework has already been successfully performed
expressing the conductance coefficients in terms of interfering electron paths
\cite{Rodol90}.

Recent experiments on magnetotransport in antidot structures unveiled the
lack of a corresponding semiclassical approach to Kubo bulk conductivity
\cite{Wei93,Nih93,Schu94}.
Antidot superlattices consist of arrays of periodically arranged holes
"drilled" through a two dimensional electron gas (2DEG). Since the lattice
constants $a$ (of typically 200--300 nm) are significantly larger than the
Fermi wavelength $\lambda_F \sim$ 50 nm the dynamics of the electrons
moving in between the repulsive antidots can
be considered to be of semiclassical nature. The combined potential of the
superlattice and a perpendicular magnetic field gives rise to a variety of
peculiar effects: The diagonal magnetoresistivity $\rho_{xx}$ exhibits
pronounced peaks due to trapping of electrons encompassing a particular number
of antidots at magnetic field strengths related to specifc values of the
normalized
cyclotron radius $R_c/a$ \cite{Wei91}.
Superimposed upon these resistivity peaks
(reflecting the {\em classical} chaotic electron dynamics of the
antidot geometry \cite{Flei92}) additional {\em quantum} resistivity
oscillations of anomalous periodicity had
been observed at sufficiently low temperature indicating phase coherence
phenomena \cite{Wei93,Nih93}. They cannot be attributed to interfering
electron waves traversing the whole device (as in the case of single junctions)
since the antidot arrays are too large to maintain phase coherence.
However, assuming that $\rho_{xx}$ reflects density of state oscillations we
were
able to describe the periodicity of the modulations observed in terms of
quantized periodic orbits in the antidot array \cite{Wei93}.
Nevertheless, a complete direct semiclassical approach to the conductivity
tensor was still missing \cite{Hack94} in spite of important work by Wilkinson
providing a semiclassical evaluation of the diagonal conductivity
in a somewhat different context
\cite{Wil87}. Such an approach is also desirable since the antidot measurements
have not yet been completely reproduced by quantum mechanical calculations
which
turn out to be rather involved \cite{Sil92}.

In this letter we derive semiclassical Kubo--type transport formulas
by stationary phase evaluation of the disorder averaged two--particle
Green functions. We obtain diagonal and Hall conductivity oscillations
in terms of sums over periodic orbit contributions with the classical actions
of the orbits determining the periodicities.
Using a model antidot potential and working at finite temperature we are
able to reproduce the amplitudes and periodicities of measured quantum
oscillations in antidot superlattices.

Within the Kubo formalism the static conductivity tensor at temperature $T$
is given by
\begin{equation}
\sigma_{ij}(T) = \frac{g_s}{V} \int \, \left( -\frac{d f}{d E}\right)
                  \, <\sigma_{ij} (E) > \, d E
\label{eq:sigma_t}
\end{equation}
where $V$ is the total area, $g_s$ accounts for the
spin degeneracy,
$f(E)=1/[1+$ $\exp((E-E_F)\beta)]$ is the Fermi function ($\beta = 1/kT$)
 and $< >$ denotes an average over weak disorder \cite{disord1}.
The diagonal and Hall conductivities $\sigma_{ij} \equiv \sigma_{ij} (E)$
 can be written as \cite{Streda82}
\begin{mathletters}
\label{eq:sigma_ij}
\begin{eqnarray}
\sigma_{xx} & = & \pi e^2 \hbar \mbox{Tr}\left\{
             \hat{v}_x \hat{\delta}(E-\hat{H})
             \hat{v}_x \hat{\delta}(E-\hat{H}) \right\}
\\
\sigma_{xy} & = &  e \frac{\partial N(E,B)}{\partial B}
                + \frac{i}{2} e^2 \hbar \mbox{Tr} \left\{
             \hat{v}_x G^+(E) \hat{v}_y \hat{\delta}(E-\hat{H}) -
             \hat{v}_x \hat{\delta}(E-\hat{H}) \hat{v}_y G^-(E) \right\} \; .
\end{eqnarray}
\end{mathletters}
Here, the $\hat{v}_i$ are the operators of the velocity components,
$G^+(E)$ $(G^-(E))$ is the advanced (retarded) Green function, $N(E)
= \int^{E} \mbox{Tr} \hat{\delta}(E'-\hat{H}) d E'$ the number of states below
$E$ and $B$ is the magnetic field.

For the semiclassical evaluation of Eq.~(\ref{eq:sigma_ij})
it is convenient to
use $\hat{\delta}(E-\hat{H}) = -[G^+(E)-G^-(E)]/2\pi i$ and to work within
position
representation. Semiclassically, the operators $\hat{v}_i$ act as classical
quantities and with $G_E^-({\bf r, r}') = {G_E^+}^\ast(
{\bf r', r})$ we can reduce Eq.~(\ref{eq:sigma_ij}) to
\begin{mathletters}
\label{eq:sigma_sc}
\begin{eqnarray}
\sigma_{xx} & = & \frac{e^2 \hbar}{2\pi} \mbox{Re} \left\{
                  \int \, d {\bf r} \, \int \, d{\bf r}'
        v_x G_{E}^+({\bf r},{\bf r}') \, v_x'\left[
{G_{E}^+}^\ast({\bf r},{\bf r}') - G_{E}^+({\bf r}',{\bf r}) \right] \right\}
\; ,
\\
\sigma_{xy} & = & e \frac{\partial N(E)}{\partial B} +
                  \frac{e^2 \hbar}{2\pi} \mbox{Re} \left\{
                  \int \, d {\bf r} \, \int \, d{\bf r}'
        v_x G_{E}^+({\bf r},{\bf r}') \, v_y'\left[
{G_{E}^+}^\ast({\bf r},{\bf r}')  - G_{E}^+({\bf r}',{\bf r}) \right]
 \right\} \; .
\end{eqnarray}
\end{mathletters}
We write the Green function as the Laplace transform of the Feynman
propagator which is semiclassically given in terms of a sum over all
classical paths $\alpha$ from ${\bf r}'$ at time $t'=0$ to ${\bf r}$ at time
$t$ \cite{BeMo72}:
\begin{equation}
G_E^+({\bf r, r}') = -\frac{1}{2\pi\hbar^2} \int_0^\infty \, dt \,
                    e^{i E t/\hbar} \, \sum_\alpha D_\alpha({\bf r',r})
                    \exp\left[\frac{i}{\hbar} W_\alpha ({\bf r}',0;{\bf r},t)
                               + m_\alpha i \frac{\pi}{2} \right] \,
                  e^{-t/2\tau_s}   \; .
\label{eq:greenfct}
\end{equation}
$W_\alpha = \int_0^t {\cal L} \,dt$ is the Hamilton principal function
$D_\alpha ({\bf r', r}) = |\mbox{det}_{ij} (-\partial {p_{r'}^\alpha}_i
 ({\bf r}',0;{\bf r},t) / \partial r_j) |^{1/2} $
contains information of the phase space in the vicinity of the trajectory
$\alpha$ and $m_\alpha$ is a topological phase. The factor $\exp(-t/2 \tau_s)$
($\tau_s$ being the elastic scattering time) results from calculating the
disorder averaged Green functions within Born
approximation \cite{Stone92}. Inserting the representation
Eq.~(\ref{eq:greenfct})
of $G_E^+$ and ${G_E^+}^\ast$ into Eq.~(\ref{eq:sigma_sc})
leads to double sums $\sum_{\alpha \bar{\alpha}}$ of products of path
contributions. We begin with the semiclassical evaluation
of the traces over the terms $v_i G_E^+({\bf r,r}') v_j'
{G_E^+}^\ast ({\bf r,r}')$ in Eq.~(\ref{eq:sigma_sc}).
The phases of the diagonal parts $(\alpha =
\bar{\alpha})$ cancel out and we obtain the {\em classical} contribution
to the conductivity in two dimensions after one time integration and a
transformation of the integral over the final coordinate
into an integral over initial momenta (by means of
$D_\alpha$ which acts as a Jacobian), i.e.,
$ \sigma^{cl}_{ij} = e^2/h^2 \int_0^\infty  dt \,
                     < v_i(t) v_j(0) >_{\bf pq}  \exp(-t/\tau_s) $ .
Here, $< >_{\bf pq}$ denotes a classical phase space average over the energy
shell.
Such a Kubo formula had been used in Ref.~\cite{Flei92} to calculate
numerically
the classical part of the conductivity for antidot arrays.

In order to take care of quantum phase coherence phenomena we have to
take into account non--diagonal $(\alpha \neq \bar{\alpha})$ parts in the sum
$\sum_{\alpha \bar{\alpha}}$. Consider, e.g.,
the contribution of Green function products of direct paths $\alpha$ (from
${\bf r}'
$ to ${\bf r}$) and paths $\bar{\alpha}$  with at least one selfcrossing as
depicted
in Fig.~\ref{fig:paths}(a). We will compute the trace integrals --- as usual
for a semiclassical approach --- by stationary phase approximation.
The stationary phase condition for the ${\bf r}$ integral requires
for the final momenta ${\bf p}^f_\alpha =  {\bf p}^f_{\bar{\alpha}}$
from which follows that the paths $\alpha$ and $\bar{\alpha}$ must
coincide between ${\bf r}'$ and ${\bf r}$ (Fig.~\ref{fig:paths}(b)).
Stationary phase approximation for the trace integral over ${\bf r}'$
selects all pairs of paths with the same initial momenta, i.e.
${\bf p}_\alpha^i =  {\bf p}_{\bar{\alpha}}^i$. Both conditions are fulfilled
for all pairs of orbits with initial and final phase space points lying on a
{\em periodic} orbit (Fig.~\ref{fig:paths}(c)).
Besides the example shown in Fig.~\ref{fig:paths}(c) we must include all pairs
of orbits with the shorter one showing an arbitrary number of full traversals
before
reaching the final point ${\bf r}$ and the longer one
having $r$ additional repetitions along the periodic orbit.
The semiclassical evaluation of the trace integrals and
one time integral follows similar lines as Gutzwiller's derivation
of the periodic orbit trace formula for the density of states
\cite{BeMo72,Gutz90,Klaus94}.
After performing finally the convolution (Eq.~(\ref{eq:sigma_t}))
with the Fermi function we obtain from the $v_i G^+_E \, v_j' \,
{G^+_E}^\ast$ terms oscillating parts of the
diagonal and Hall conductivity as  sums over contributions from all primitive
periodic orbits (po) of the classical system and their repetitions $r$:

\begin{mathletters}
\label{eq:sigma_end}
\begin{eqnarray}
\label{eq:sigma_xx}
\sigma_{xx}^{osc}(E_F,B;T) & = & \frac{2g_s}{V}\frac{e^2}{h}\;
                        \sum_{po} \,{\cal C}_{xx}^{po}
                        \sum_{r=1}^\infty \, \frac{R_r (\beta)
\, e^{-r T^{po}/(2\tau_s)} }{\left|
\mbox{det}({\bf M}_r^{po} - {\bf 1})\right|^{1/2} }
\cos\left[r(S^{po}/\hbar -\mu^{po} \pi/2)\right]  \; ,
\\
\sigma_{xy}^{osc}(E_F,B;T) & = & \frac{2g_s}{V}\frac{e^2}{h}\; \sum_{po} \,
                  \sum_{r=1}^\infty \, \left(\frac{r}{e}\frac{\partial S^{po}}{
                         \partial B} + {\cal C}_{xy}^{po}\right)\,
\frac{R_r (\beta) \, e^{-r T^{po}/(2\tau_s)}
}{\left|\mbox{det} ({\bf M}^{po}_r - {\bf 1)}\right|^{1/2} }
\,   \cos\left[r(S^{po}/\hbar -\mu^{po} \pi/2)\right]  \; .
\end{eqnarray}
\end{mathletters}
In Eq.~(\ref{eq:sigma_end})
$S^{po}(E,B) = \oint_{po} {\bf p}d{\bf q}$ is the classical
action, $T^{po}$ the period and $\mu^{po}$ the Morse index of a
periodic orbit. The monodromy matrix ${\bf M}_r^{po}$ measures its classical
stability \cite{Gutz90}.
\begin{equation}
{\cal C}_{ij}^{po} =
\int_0^\infty d t \, e ^{-t/\tau_s} \; \int_0^{T^{po}} d\tau \,
v_i(\tau) \, v_j(t+\tau)
\label{eq:cij}
\end{equation}
 are velocity--correlation functions for motion along the periodic orbits.
The function
\begin{equation}
R_r (\beta) = \frac{r T^{po}/\tau_\beta}{\sinh
 (r T^{po}/\tau_\beta)}  \hspace{1cm} ; \hspace{1cm} \tau_\beta =
   \frac{\hbar \beta}{\pi}
\label{eq:sinh}
\end{equation}
gives rise to temperature damping which is exponential for orbits
with $T^{po} \gg \tau_\beta$.

Stationary phase evaluation of the trace integrals in Eq.~(\ref{eq:sigma_ij})
for
the terms $v_i G^+_E({\bf r, r}') v_j' G^+_E({\bf r}',{\bf r})$ representing
recurring paths from ${\bf r}$ via ${\bf r}'$ leads also to periodic orbits
(see Fig.~\ref{fig:paths}(d--f)). However, the corresponding velocity
correlation functions vanish and therefore $v_i G^+ v_j' G^+$ terms do not
contribute to $\sigma_{ij}$ semiclassically.

Up to the correlation functions ${\cal C}^{po}_{xx}$ the trace formula
Eq.~(\ref{eq:sigma_xx}) for $\sigma_{xx}$ is exactly the same as Gutzwiller's
trace formula for the density of states. $\sigma_{ij}$ quantum oscillations
are semiclassically related to interference (due to the phase differences
$\sim r S_{po}/\hbar$) between pairs of paths along a periodic orbit differing
by $r$ in their number of full traversals. This approach therefore requires
only
phase coherence on the length scale of a periodic orbit and {\em not}
throughout the entire system.

In the following we will apply our results to magnetotransport in 2DEG's.
In the unmodulated case the electrons follow cyclotron orbits with action
$S_{cyc} = 2\pi E_F/ \omega$ and frequency $\omega = e B/m^\ast$. Then
the classical amplitudes\cite{Klaus2} and velocity correlation functions
in the trace formulas (\ref{eq:sigma_end}) can be calculated analytically.
For the diagonal conductivity we obtain, e.g.,
\begin{equation}
\sigma_{xx} =
            \frac{n_s e^2 \tau_s}{m^\ast}
            \frac{1}{1+(\omega \tau_s)^2}
              \, \left[1+2\sum_{r=1}^\infty (-1)^r\,
                R_r(\beta) \cos\left(r\frac{
               2\pi E_F}{\hbar\omega}\right) \; \exp \left(\frac{-r \pi}{
              \omega\tau_s} \right) \right]
\label{eq:sdh}
\end{equation}
($n_s$ being the carrier density) which represents a semiclassical
approximation
of the Shubnikov de Haas oscillations and
 coincides with the quantum mechanical result for a constant scattering
time $\tau_s$ \cite{Ando82}. In the case of a high mobility 2DEG
($\omega \tau_s \gg 1$) with a distinct Landau level structure
the use of a density of state dependent
scattering time  is usually required in order
to obtain the correct conductivity amplitudes.
However, in the antidot arrays which we consider in the following
the periodic superlattice potential strongly mixes the Landau levels
\cite{Sil92} and therefore justifies the use of an energy independent $\tau_s$.
The inset of Fig.~\ref{fig2}(a) depicts the measured magnetoresistivity
$\rho_{xx}(B)$ of an antidot array for two temperatures T$=$4.7 K
and 0.4 K. While the gross structure of the broad (classical)
maximum at $2R_c(B)=a$ persists up to
temperatures of 60 K the (quantum) oscillations at T$=$0.4 K on top of the
maximum disappear with increasing temperature (at T$=$4.7 K).

In order to
show that the quantum oscillations at T$=$0.4 K are related to
$\sigma_{xx}^{osc}(B)$ we compare the experimental
difference signal $\Delta \sigma_{xx} = \sigma_{xx}$(T=0.4K)$-\sigma_{xx}$
(T=4.7K) shown in Fig.~\ref{fig2}(a) with our semiclassical predictions for
$\sigma_{xx}^{osc}(B)$. As a model of the antidot lattice we use the potential
$V(x,y)=V_0 |\sin(\pi x/a) \sin(\pi y/a)|^\beta$ which has already proved
useful to describe magnetotransport in antidot arrays \cite{Wei93,Flei92}.
$V_0$ is determined through the normalized width $d/a$ of the antidots
at the Fermi energy and $\beta$ (governing the steepness of the antidots)
remains as the only free parameter. Here we use $d/a =$ 0.5 and $\beta=2.3$.

In order to calculate numerically $\sigma_{xx}^{osc}(B)$ it is convenient
\cite{WRT92}
to expand the prefactors in Eq.~(\ref{eq:sigma_xx}) into geometrical series
and to perform the sum over $r$ to get the following form
\begin{equation}
\sigma_{xx}^{osc}(E_F,B;T) = \frac{4g_s}{V}\frac{e^2}{h}\; Re\left\{
                        \sum_{po} \,{\cal C}_{xx}^{po}
\frac{T^{po}}{\tau_\beta}
                        \sum_{k,l=0}^\infty \,
  \frac{t_{po}^{(k,l)}}{(1-t^{(k,l)}_{po})^2} \right\}
\label{eq:zeta}
\end{equation}
with
\begin{equation}
t^{(k,l)}_{po} = (\pm1)^k \; \exp\left[i\left(\frac{S^{po}}{\hbar}-\frac{
                  \pi}{2}\mu^{po}\right)-\left(k+\frac{1}{2}\right)
\lambda^{po}
                - (2l+1)\frac{T^{po}}{\tau_\beta}-\frac{T^{po}}{2\tau_s}
                      \right]
\label{eq:t_po}
\end{equation}
Here, $\lambda^{po} = i\gamma^{po} $, $\gamma^{po} $ (real) being the winding
number, $k$
a semiclassical quantum number for a stable periodic orbit and $\lambda^{po} >
0$
(real) being the Liapunov exponent in case of an unstable periodic orbit
\cite{WRT92}. The ``$-$'' sign in Eq.~(\ref{eq:t_po}) applies to
unstable inverse hyperbolic orbits; otherwise the ``$+$'' sign holds.
A correct application of Eq.~(\ref{eq:zeta}) requires a detailed knowledge
of the phase space structure of the underlying classical motion in the
potential of the antidots and the magnetic field. It turns out
that the major part of the classical phase space is chaotic with a few
islands of stable motion imbedded which vanish with decreasing magnetic
field \cite{Klaus3}. Under the experimental conditions (T $=$ 0.4 K
and $\omega \tau_s \approx 2$ at $2R_c = a$ \cite{exp}) only the shortest
periodic orbits contribute
significantly to $\sigma_{xx}^{osc}$ since the terms from longer orbits
($T^{po} > \tau_s$ or $T^{po} > \tau_\beta$) are exponentially small. For our
calculations we take into account the seven fundamental
periodic orbits shown in Fig.~\ref{fig2}(d).

Our result for $\sigma_{xx}^{osc}$ (at T$=$0.4 K) is shown as the full line
 in Fig.~\ref{fig2}(b): $\sigma_{xx}^{osc}(B)$ oscillates
with the same frequency as the measured conductivity.  The period of the
oscillations is nearly constant with respect to $B$ in contrast to the $1/B$
periodic behaviour of ordinary Shubnikov de Haas
oscillations (Eq.~(\ref{eq:sdh})) which are shown in Fig.~\ref{fig2}(c)
for an unmodulated 2DEG under the same conditions. The B periodic behaviour
of quantum oscillations in certain antidot arrays can be related to
the mechanism that flux enclosing periodic orbits between antidots
cannot expand with decreasing $B$ as in the case of free cyclotron motion
\cite{Wei93}. Having in mind that the antidot potentials are not precisely
known
also the semiclassical amplitudes  are on the whole in
agreement with the measured curve of $\Delta \sigma_{xx}$.
They show --- as in the experiment --- an irregular behaviour  resulting from
interference effects between different periodic orbits and from the
magnetic field dependence of the classical orbit parameter entering
Eq.~(\ref{eq:zeta}).
The semiclassical curves show also the correct temperature dependence:
Due to the temperature damping $R_r(\beta)$ (Eq.~(\ref{eq:sinh})) the
periodic orbit oscillations decrease for T = 2.5 K (dotted line in
Fig.~\ref{fig2}(b))
and nearly disappear at T = 4.7 K (dashed curve) as in the experiment
(dashed curve in the inset of Fig.~\ref{fig2}(a)).
Fig. \ref{fig2}(e) illustrates the $\omega \tau_s$ dependence of
$\sigma^{osc}_{xx}(B)$
for three values $\omega \tau_s$ = 1, 2 and 5 showing that the result does not
strongly
depend on the $\tau_s$ chosen.

In summary we derived semiclassical formulas for the magnetoconductivity
tensor within the Kubo approach and applied them to quantum transport
in antidot lattices.
However, it remains to understand the semiclassical relation between the Kubo
formalism
leading to {\em periodic} orbit contributions and the semiclassical Landauer
B\"uttiker approach based on interference effects due to
{\em non--closed} classical trajectories since the two corresponding quantum
mechanical approaches are known to be equivalent under certain conditions
\cite{Bar89}.

I would like to thank R.~A.~Jalabert, P.~Leb{\oe}uf, H.~Silberbauer,
M.~Suhrke, D.~Ullmo, D.~Weiss and D.~Wintgen for helpful discussions
and especially D.~Weiss for providing the experimental data.
I gratefully acknowledge financial support by the Humboldt foundation.


\begin{figure}
\caption{
Typical semiclassical paths $\alpha, \bar{\alpha}$ contributing to
(a) $v_i G_E^+({\bf r, r}') v_j' {G^+_E}^\ast ({\bf r, r}')$ and
(d) $v_i G_E^+({\bf r, r}') v_j' G^+_E ({\bf r}',{\bf r})$.
The stationary phase condition at ${\bf r}$ selects paths with
${\bf p}_\alpha({\bf r}) = {\bf p}_{\bar{\alpha}}({\bf r}$) (b,e). Paths along
periodic orbits result from the stationary phase condition
${\bf p}_\alpha({\bf r}') = {\bf p}_{\bar{\alpha}}({\bf r}'$) at ${\bf r}'$
(c,f).
}
\label{fig:paths}
\end{figure}

\begin{figure}
\caption{
(a) Oscillatory part $\protect \Delta \sigma_{xx} = \sigma_{xx}(0.4K) -
\sigma_{xx}(4.7K)$ of the experimental \protect\cite{Wei93}
diagonal conductivity ($\sigma_{xx} = \rho_{xx}/ (\rho_{xx}^2+\rho_{xy}^2)$)
 of an antidot array as a function of an applied magnetic field.
Inset: Measured resistivity $\rho_{xx}$ at T $=$ 0.4 K
(full line) and T $=$ 4.7 K (dashed line).
(b) Semiclassically calculated oscillatory part $\sigma_{xx}^{osc}$
(from Eq. \protect \ref{eq:zeta}) for three different temperatures T $=$ 0.4 K
(full line), T $=$ 2.5 K (dotted line) and T $=$ 4.7 K (dashed line).
(c) Semiclassical Shubnikov de Haas oscillations (Eq.~\protect\ref{eq:sdh})
for an unmodulated 2DEG under the same conditions as in (b).
(d) fundamental periodic orbits in a model antidot potential
which enter the semiclassical calculation.
(e) $\protect \sigma^{osc}_{xx}$ at T $=$ 0.4 K for $\protect \omega \tau_s =
5$ (dotted line), 2 (full line) and 1 (dashed line). }
\label{fig2}
\end{figure}

\begin{references}
\bibitem{disord1} for an antidot array an additional ensemble average over
                    variations in the periodic potentials
                   in different parts of the array
                   has to be performed. However, for the experiments addressed,
                   these variations are negligible (D.~Weiss, private
                   communication).
\bibitem{Klaus3} The trace formulas (\ref{eq:sigma_end}) do not rigorously
apply
                 to systems with mixed phase space. However, since the antidot
system
                 is nearly chaotic with a few distinct stable islands our
procedure
                 to calculate $\sigma^{osc}_{ij}$ can be regarded as treating
regular
                 and chaotic phase space parts separately.
\bibitem{exp}    The effective value of $\omega \tau_s$ is reduced
                  with respect to the unmodulated 2DEG
                 due to the presence  of the antidot potentials and can only be
estimated.
\end{references}
\end{document}